\documentclass[aps,prb,superscriptaddress,reprint,floatfix,balancelastpage,showpacs]{revtex4-1}
\usepackage{graphicx,amsmath,amssymb,bm,color,engord,nicefrac,placeins}
\usepackage[caption=false,labelformat=empty,position=t]{subfig}
\usepackage[usenames,dvipsnames]{xcolor}
\usepackage[colorlinks,colorlinks,urlcolor=Magenta,citecolor=Magenta,linkcolor=black]{hyperref}
\begin{document}

\title{Fractionally charged impurity states of a fractional quantum Hall system}
\author{Kelly R. Patton}
\email[\hspace{-1.4mm}]{kpatton@physast.uga.edu}
\affiliation{Seoul National University, Department of Physics and Astronomy\\ Center of Theoretical Physics, 151-747 Seoul, South Korea }
\author{Michael R. Geller}
\affiliation{Department of Physics and Astronomy, University of Georgia, Athens, GA 30602}
\date{\today}
\begin{abstract}
The single-particle spectral function for an incompressible fractional quantum Hall state in the presence of a scalar short-ranged attractive impurity potential is calculated via exact diagonalization within the spherical geometry.  In contrast to the noninteracting case, where only a single bound state below the lowest Landau level forms, electron-electron interactions strongly renormalize the impurity potential, effectively giving it a finite range,  which can support  many quasi-bound states (long-lived resonances). Averaging the  spectral weights of the quasi-bound states and extrapolating to the thermodynamic limit, for filling factor $\nu=1/3$ we find  evidence consistent with localized fractionally charged $e/3$ quasiparticles. For $\nu=2/5$, the results are slightly more ambiguous, due to finite size effects and possible bunching of Laughlin-quasiparticles. 
\end{abstract}
\maketitle

\section{Introduction}
\label{sec: introduction}

The ubiquitous presence of impurities in condensed matter systems is often a contentious problem.  But they are also responsible for such intriguing  phenomena as Anderson localization and the Kondo effect.  While impurities and their effects are often a concern or subject of experiments,   they can also be used as an experimental tool.  For instance, by probing the scattering or bound states induced by a single localized impurity, information about the  underlying microscopic---as opposed to thermodynamic---properties of the impurity-free bulk can be obtained.  A striking example of this is in high-temperature superconducting systems, where tunneling experiments have been able to measure the local density of states near an impurity atom, in the superconducing state. \cite{PanNature00,BalastskyRMP06}  The spatial structure of which is related to the symmetry of the superconducting order parameter, or Cooper-pair wave function of the bulk, e.g., $s$- or $d$-wave. Similarly, a fractional quantum Hall (FQH) system can  support exotic quasiparticles that  carry a fractional charge.  These quasiparticles, predicted by Laughlin,\cite{LaughlinPRL83} were first observed in the nonequilibrium  shot-noise of the current carrying  FQH edge states.\cite{RdePicciottoNature97} More recently, experimental evidence of fractionally charged quasiparticles occurring in the bulk\cite{MartinScience04} has been found, as well as through numerical simulations. \cite{RezayiPRB85,TsiverPRL06,HuPRB08}

On the theoretical side, the effects of impurities on FQH systems has been a topic of great interest.   Prior works have mostly focused on the effects  of a single localized  scalar or magnetic potential on the local density. \cite{RezayiPRB85,ZhangPRB85,AristonePRB93,VybornyArxiv07}   Here, we report on another interesting aspect, which involves  probing the impurity-induced bound states in a FQH system to test for the existence of localized fractionally charged Laughlin-quasiparticles.   To this end, we numerically calculated the spectral function of a FQH droplet by exact diagonalization in the presence of an attractive impurity potential and extracted the spectral weights of the resulting bound states.  In principle the bound-state spectral weight(s)   correspond to the fraction of a bare electron in the bound state, i.e., the fractional charge. 

 In a noninteracting system the  spectral weight of a bound state $Z_{\rm b}$ is unity.  For a Landau-Fermi liquid $Z_{\rm b}\simeq Z$, where $Z$ is the so-called wave-function renormalization, or quasiparticle amplitude, which  roughly corresponds to the fraction of a bare electron that remains in a quasiparticle.  Because an incompressible ground state of a FQH system is believed to support fractionally charged Laughlin quasiparticles, the presence of an attractive impurity potential could bind one or more of these quasiparticles into a localized state.  
 
Furthermore, Jain's highly successful composite fermion theory \cite{CompositeFermionsJainBook,JainPRL05} and the elastic model of Conti and Vignale\cite{ContiJPhysCondMat98,VignalePRB06}  predict a Fermi-liquid-like spectral peak at the chemical potential that corresponds to a single-particle excitation of a bound complex of multiple composite fermions, which re-forms an electron quasiparticle.  The  electron-quasiparticle peak has not been observed in tunneling experiments\cite{,EisensteinPRL92,BoebingerPRB93,EisensteinPRL95,EisensteinSolidSatcom09} or in exact diagonalization studies of finite-sized systems,\cite{HePRL93} including the present one.  These non-Laughlin-quasiparticles, if present, could also be bound to an impurity potential.  A signature of this would be a bound-state spectral weight that approaches unity.  For the largest system sizes studied in the present work, we do find such a state, but, owing to finite size effects, its identification as an electron-quasiparticle remains tenuous. 

In the following section, we outline how impurity bound states and energies can be identified within standard diagrammatic many-body theory for the single-particle Green's function, with and without electron-electron interactions, using the $T$-matrix formalism.  After which,  details are given in Secs.~\ref{sec: FQHE on a Haldane sphere} and \ref{sec: Spectral function}  for the exact numerical calculations of the Green's function and relevant spectral function for a finite-sized fractional quantum Hall system, in the spherical geometry.  Finally,  results and conclusions are discussed in Sec.~\ref{sec: Results}. 

\section{Determination of Impurity bound states}
\label{sec: T-matrix etc}

The location of the possibly complex poles, in frequency space,  of the single-particle Green's function determines the energies and lifetimes of the single-particle excitations  of the system. \cite{AGD} For example, in a noninteracting impurity-free system, the Fourier transformed zero-temperature (retarded)   Green's function is $(\hbar=1)$
\begin{equation}
\label{eq: noninteracting Green's function}
G^{}_{0}({\bm r},\omega)=\lim_{\eta\to 0^{+}}\frac{1}{\sf V}\sum_{\bm k}\frac{e^{i{\bm k}\cdot{\bm r}}}{\omega-(\epsilon_{\bm k}-\mu)+i\eta},
\end{equation}
where $\sf V$ is the system volume, $\mu$ the chemical potential, and $\epsilon^{}_{\bm k}$ is the dispersion. 
As one  expects, the poles of $G^{}_{0}(\omega)$ occur when $\omega=\epsilon_{\bm k}-\mu$.   In the presence of an attractive impurity potential, additional poles can form.  These indicate the  impurity-induced bound states.  For a static impurity potential $V({\bm r})$, and neglecting electron-electron interactions for the moment, the Dyson equation for the time-ordered Green's function is 
\begin{align}
\label{eq: clean Dyson equation}
G^{}({\bm r},{\bm r}'&,\omega)=G^{}_{0}({\bm r},{\bm r}',\omega)\nonumber\\&+\int d{\bm r}_{1}d{\bm r}_{2}\, G^{}_{0}({\bm r},{\bm r}^{}_{1},\omega)\Sigma({\bm r}_{1},{\bm r}_{2},\omega)G({\bm r}_{2},{\bm r}',\omega),
\end{align}
where the self-energy $\Sigma({\bm r},{\bm r}',\omega)=V({\bm r})\delta({\bm r}-{\bm r}')$.   Any additional poles of $G(\omega)$ that are related to the impurity are more easily determined  by rewriting  Dyson's equation \eqref{eq: clean Dyson equation} into a $T$-matrix equation.  By iterating \eqref{eq: clean Dyson equation} the Green's function can also be expressed as
\begin{align}
\label{eq: clean T-matrix equation for G}
G^{}({\bm r},{\bm r}'&,\omega)=G^{}_{0}({\bm r},{\bm r}',\omega)\nonumber\\&+\int d{\bm r}_{1}d{\bm r}_{2}\, G^{}_{0}({\bm r},{\bm r}^{}_{1},\omega)T({\bm r}_{1},{\bm r}_{2},\omega)G_{0}({\bm r}_{2},{\bm r}',\omega),
\end{align}
where the so-called $T$-matrix is given by
\begin{align}
\label{eq: clean T-matrix equation}
T({\bm r},{\bm r}',&\,\omega)=\Sigma({\bm r},{\bm r}',\omega)\nonumber\\ &+\int d{\bm r}_{1}d{\bm r}_{2}\Sigma({\bm r},{\bm r}_{1},\omega)G_{0}({\bm r}_{1},{\bm r}_{2},\omega)T({\bm r}_{2},{\bm r}',\omega).
\end{align}
As the poles of the clean system are given by $G_{0}(\omega)$, from Eq.~\eqref{eq: clean T-matrix equation for G} any additional ones, related to the impurity,  must be determined by poles of the $T$-matrix.  
 
 In one-dimension and for a delta-function impurity potential, the $T$-matrix  equation can be straightforwardly solved. Its single pole $\omega_{\rm b}$, below the bottom of the band, corresponds to the well-known solution of  Schr\"odinger's equation for the bound-state energy of an attractive delta-function potential.  Near the bound-state energy, the local density of states $N({\bm r},\omega)=-\frac{1}{\pi}{\rm Im}\,G^{}({\bm r},{\bm r},\omega)$ can be written as
 \begin{equation}
 N({\bm r},\omega\to\omega_{\rm b})=|\psi_{\rm b}({\bm r})|^{2}\delta(\omega-\omega_{\rm b}),
 \end{equation}
 where $\psi_{\rm b}({\bm r})$ is the normalized  bound-state wave function.  The spectral weight $Z_{\rm b}$ of the bound state is then 
 \begin{equation}
 Z_{\rm b}=\int d{\bm r}\int\limits_{\omega^{}_{\rm b}-0^{+}}^{\omega^{}_{{\rm b}}+0^{+}}d\omega\,N({\bm r},\omega)=1. 
 \end{equation}
 Although in two- and three-dimensions a delta-impurity potential  must be treated more carefully, bound states in noninteracting  quantum Hall systems have been extensively studied. \cite{PrangePRB81,KhalilovJPhysA07,FernanoAJP91,CavalcantiJPhysAMath98}
 
As electron-electron interactions dominate in the fractional quantum Hall regime, they  must also be included in the calculation of the electron Green's function, along with the impurity potential. The $T$-matrix formulation of the impurity problem can be generalized to a interacting system.\cite{ZieglerPRB96} In general, interactions lead to a finite lifetime of bound states producing  a resonance, or quasi-bound state.  Additionally, as  shown in the following,  interactions renormalize the impurity potential.  This renormalized or effective potential can support additional quasi-bound states. 
 
For a system with electron-electron interactions, for example Coulomb, Dyson's equation \eqref{eq: clean Dyson equation} retains the same form, but now  the self-energy $\Sigma(\omega)$ contains both  electron-electron correlations and impurity scattering diagrams.  These two contributions can be formally separated by writing $\Sigma(\omega)=\Sigma_{U}(\omega)+\Sigma_{UV}(\omega)$, where $\Sigma_{U}(\omega)$ contains all irreducible diagrams that involve only the Coulomb interaction between electrons, and  $\Sigma_{UV}(\omega)$ contains all diagrams with at least one impurity potential.  $\Sigma_{UV}(\omega)$ accounts for the bare impurity potential and terms involving  combinations of both the impurity and Coulomb interaction to all orders.  Dyson's equation can then be formally rewritten as 
\begin{align}
\label{eq: interacting impurity Dyson equation}
G^{}({\bm r},&{\bm r}',\,\omega)=G^{}_{U}({\bm r},{\bm r}',\omega)\nonumber\\&+\int d{\bm r}_{1}d{\bm r}_{2}\, G^{}_{U}({\bm r},{\bm r}^{}_{1},\omega)\Sigma^{}_{UV}({\bm r}_{1},{\bm r}_{2},\omega)G_{}({\bm r}_{2},{\bm r}',\omega),
\end{align}
where $G^{}_{U}(\omega)$ is the exact interacting Green's function in the absence of the impurity.  In analogy with the noninteracting case, Eq.~\eqref{eq: clean Dyson equation}, the diagonal matrix elements (in position space) of $\Sigma_{UV}({\bm r},{\bm r}',\omega)$ can then be identified as an effective or renormalized impurity potential  $V_{\rm eff}({\bm r},\omega)$.  In general even if $V_{}({\bm r})$ is short-ranged, $V_{\rm eff}({\bm r},\omega)$ is not.  Furthermore, a generalized $T$-matrix equation can also be found from \eqref{eq: interacting impurity Dyson equation} and can be formally expressed by
\begin{align}
\label{eq: interacting T-matrix}
T({\bm r},\,&{\bm r}',\omega)=\Sigma_{UV}({\bm r},{\bm r}',\omega)\nonumber\\&+\int d{\bm r}_{1}d{\bm r}_{2}\,\Sigma_{UV}({\bm r},{\bm r}_{1},\omega)G_{U}({\bm r}_{1},{\bm r}_{2},\omega)T({\bm r}_{2},{\bm r}',\omega).
\end{align}
 The poles of which determine the quasi-bound states of an interacting system.

\section{FQHE on a Haldane sphere}
\label{sec: FQHE on a Haldane sphere}
Haldane was the first to introduce the spherical geometry to study boundaryless  finite-sized fractional quantum Hall systems.\cite{HaldanePFL83} The electrons are confined to the surface of a two-dimensional sphere of radius $R$. The quantizing magnetic field ${\bm B}$ is produced by a Dirac magnetic monopole located at the center of the sphere;
\begin{equation}
\label{eq: magnetic monopole field}
{\bm B}({\bm r})=\frac{2Q\Phi_{0}}{4\pi r^{2}_{}}\hat{\bm r},
\end{equation}
where  $Q$ is the so-called monopole strength, which can be a positive or negative integer or half-integer, and $\Phi_{0}=hc/e$ is the flux quantum.  The total magnetic flux through the surface of the sphere is then $2|Q|\Phi_{0}$, i.e., $2|Q|\Phi_{0}=4\pi R^{2}|{\bm B}|$;  thus, in units of $\hbar=1$ and $\ell=\sqrt{c/(e|\bm{B}|)}$ the sphere's radius is $R=\ell \sqrt{|Q|}$.  
Various vector potentials ${\bm A}$ such that $\nabla\times{\bm A}={\bm B}$ which differ by the  location and number of Dirac strings can be commonly found in the literature.  Here, we use  
\begin{equation}
\label{eq: monopole vector potential}
{\bm A}({\bm r})=-\frac{c\,Q}{er}\cot(\theta)\hat{\boldsymbol{\phi}}. 
\end{equation}

The noninteracting first quantized Hamiltonian for an electron with charge $-e$ confined to the surface of the sphere is then taken to be
\begin{equation}
\label{eq: noninteracting Hamiltonian}
H^{}_{0}=\frac{1}{2m}\Big[-\frac{i}{R}\nabla^{}_{\Omega}+\frac{e}{c}{\bm A}\Big]^{2},
\end{equation}
where $\nabla_{\Omega}=\hat{\boldsymbol{\theta}}\partial_{\theta}+\frac{1}{\sin(\theta)}\hat{\boldsymbol{\phi}}\partial_{\phi}$.  Defining the  gauge-invariant orbital angular momentum
\begin{equation}
\label{eq: gauge invariant orbital momentum}
\boldsymbol{\Lambda}=R\hat{\bm r}\times \Big[-\frac{i}{R}\nabla^{}_{\Omega}+\frac{e}{c}{\bm A}\Big],
\end{equation}
the Hamiltonian \eqref{eq: noninteracting Hamiltonian} can then be expressed as 
\begin{equation}
\label{eq: noninteracting Hamiltonian in terms of Lambda}
H^{}_{0}=\frac{\boldsymbol{\Lambda}^{2}}{2mR^{2}}. 
\end{equation}
The one electron normalized  eigenfunctions of \eqref{eq: noninteracting Hamiltonian in terms of Lambda} are given by the so-called monopole, or spin-weighted, spherical harmonics\footnote{See Ref. \onlinecite{CompositeFermionsJainBook} and references therein for further details.}
\begin{align}
\label{eq: monopole spherical harmonics}
_{Q}Y_{lm}(\boldsymbol{\Omega})&=N_{Qlm}2^{-m}(1-x)^{\frac{m-Q}{2}}(1+x)^{\frac{m+Q}{2}}\nonumber\\& \times P^{m-Q,m+Q}_{l-m}(x)e^{imQ},
\end{align}
with $x=\cos(\theta)$,
\begin{equation}
\label{eq: monopole spherical harmonics norm}
N_{Qlm}=\left(\frac{2l+1}{4\pi}\frac{(l-m)!}{(l-Q)!}\frac{(l+m)!}{(l+Q)!}\right)^{1/2},
\end{equation}
and $P^{\alpha,\beta}_{n}(x)$ are the Jacobi polynomials defined by
\begin{equation}
\label{eq: Jacobi polynomials }
P^{\alpha,\beta}_{n}(x)=\frac{1}{2^{n}}\sum_{j=0}^{n}\binom{n+\alpha}{j}\binom{n+\beta}{n-j}(x-1)^{n-j}(x+1)^{j}. 
\end{equation}
The allowed values  of $l$ and $m$ are
\begin{align}
\label{eq: allowed quantum numbers}
l&=|Q|,|Q|+1,\ldots\nonumber\\&
m=-l,-l+1,\ldots, l.
\end{align} 
The associated  energy eigenvalues are
\begin{equation}
\label{eq: single-particle energy levels}
\epsilon_{l}=\frac{l(l+1)-Q^{2}}{2|Q|}\omega_{\rm c},
\end{equation}
where $\omega_{\rm c}=e|{\bm B}|/(mc)$.  In  the lowest Landau level (LLL) $l=|Q|$, the eigenfunctions \eqref{eq: monopole spherical harmonics} simplify to 
\begin{align}
\label{eq: LLL eigenfunctions}
_{Q}Y_{Qm}(\boldsymbol{\Omega})&=(-1)^{Q-m}\left[\frac{2Q+1}{4\pi}\binom{2Q}{Q-m}\right]^{1/2}\nonumber\\ &\times \cos^{Q+m}(\theta/2)\sin^{Q-m}(\theta/2)e^{imQ}.
\end{align}
Because of the compact geometry of the sphere, the degeneracy of the LLL is finite. The single-particle Hilbert space is span by only $2|Q|+1$ states.  

In the LLL approximation the kinetic energy can be neglected; thus, for $N$ particles interacting through the Coulomb potential $U$  the LLL Hamiltonian is simply 
\begin{align}
\label{eq: Coulomb Hamiltonian}
H_{\rm LLL}=\frac{1}{2}\sum^{N}_{i\neq j}U(\boldsymbol{\Omega}_{i}-\boldsymbol{\Omega}_{j})=\frac{e^{2}}{2R}\sum_{i\neq j}^{N}\frac{1}{|\boldsymbol{\Omega}_{i}-\boldsymbol{\Omega}_{j}|},
\end{align}
where the chord distance between the particles on the sphere has been used.  The two-body matrix elements of the  Coulomb interaction in the LLL Hilbert space are given by\cite{FanoPRB86}
\begin{align}
\label{eq: two-body Coulomb matrix elements}
&\langle Q m_{1}, Qm_{2}|U|Qm'_{1},Qm'_{2}\rangle=\nonumber\\&\frac{e^{2}}{R}\sum_{L=0}^{2Q}\sum_{M=-L}^{L}U^{(Q)}_{L}\langle Qm_{1},Qm_{2}|LM\rangle\langle Qm'_{1},Qm'_{2}|LM\rangle,
\end{align}
where $\langle Qm_{1},Qm_{2}|LM\rangle$ are Clebsch-Gordan coefficients and
\begin{equation}
U^{(Q)}_{L}=2\frac{\binom{2Q-2L}{2Q-L}\binom{4Q+2L+2}{2Q+L+1}}{\binom{4Q+2}{2Q+1}^{2}}. 
\end{equation}

A short-ranged attractive  impurity potential with strength $g <0$ located at the north pole of the sphere  can be written as 
\begin{equation}
\label{eq: impurity potential}
V_{}(\boldsymbol{\Omega})=\frac{g\,\delta(\theta)}{2\pi \sin(\theta)}. 
\end{equation}
The impurity matrix elements in the LLL Hilbert space are  then
\begin{align}
\label{eq: matrix elements of impurity}
\langle Qm|V|Qm'\rangle&=\frac{g (2Q+1)}{4\pi}\delta_{Q,m}\delta_{Q,m'}\nonumber\\&\equiv\lambda \delta_{Q,m}\delta_{Q,m'}.
\end{align}

In general the relationship between the number of particles $N$, the total number of magnetic flux quanta $N_{\Phi}$, and the filling factor $\nu$ of the FQH state depends on the topology of the system, and is given by
\begin{equation}
\label{eq: defining the filling factor}
N_{\Phi}=N\nu^{-1}-S,
\end{equation}
where $S$ is a topological quantum number commonly called the shift.\cite{QHEbook,dAmbrumenilPRB89,WenPRL92,BanerjeePRB98}   For a sphere $N_{\Phi}=2Q$, and for filling factors $\nu=p/(2p+1)$ with $p\in \mathbb{N}$, the shift is $S=2+p$.

For a given filling factor and number of electrons, the corresponding  monopole strength can  be determined and the $N$-particle LLL Hilbert space basis constructed using the single-particle states, Eq.~\eqref{eq: LLL eigenfunctions}.  The matrix elements of the $N$-particle Hamiltonian \eqref{eq: Coulomb Hamiltonian}, with or without an impurity \eqref{eq: impurity potential}, in the $N$-particle basis can  be obtained using Slater-Condon rules\cite{SzaboModchem} along with the one- and two-body matrix elements  \eqref{eq: matrix elements of impurity} and \eqref{eq: two-body Coulomb matrix elements}.   Numerical diagonalization of the resulting Hamiltonian then allows one to calculate the Green's function, as shown in the following section.  An exact full diagonalization calculation, as apposed to iterative methods, such as Lanczos, for the single-particle Green's function is preformed.  This restricts our calculations to smaller systems sizes, but provides highly accurate spectral weights for all states, which is required. 
  
\section{Spectral function}
\label{sec: Spectral function}
At zero temperature the retarded single-particle  Green's function for the angular coordinates ${\boldsymbol{\Omega}}=(\theta,\phi)$  is defined as
\begin{align}
\label{eq: ret Green's function}
G^{\rm r}_{}(\boldsymbol{\Omega},\boldsymbol{\Omega}',t)&=-i\Theta(t)\langle\psi^{\rm g}_{N}|\{\hat{\Psi}(\boldsymbol{\Omega},t),\hat{\Psi}^{\dagger}(\boldsymbol{\Omega}',0)\}|\psi^{\rm g}_{N}\rangle,
\end{align}
where $|\psi^{\rm g}_{N}\rangle$ is the exact ground state of an interacting  $N$-particle system, and $\hat{\Psi}^{(\dagger)}(\boldsymbol{\Omega})$ are the second quantized fermionic  annihilation (creation) operators.  Even in the presence of the impurity potential \eqref{eq: impurity potential},  $L_{z}$ remains a conserved quantity. Thus, in the LLL approximation 
\begin{align}
\label{eq: LLL ret Green's function}
G^{\rm r}_{}(\boldsymbol{\Omega},\boldsymbol{\Omega}',t)&\approx -i\Theta(t)\sum_{m=-Q}^{Q} \phantom{}_{Q}Y_{Qm}(\boldsymbol{\Omega}) \phantom{}_{Q}Y^{*}_{Qm}(\boldsymbol{\Omega}')\nonumber\\&\times \langle\psi^{\rm g}_{N}|\{\hat{a}^{}_{m}(t),\hat{a}^{\dagger}_{m}(0)\}|\psi^{\rm g}_{N}\rangle\nonumber\\&=\sum_{m=-Q}^{Q} \phantom{}_{Q}Y_{Qm}(\boldsymbol{\Omega}) \phantom{}_{Q}Y^{*}_{Qm}(\boldsymbol{\Omega}')G^{\rm r}_{m}(t)
\end{align}
The $L_{z}$-resolved  Green's function  $G^{\rm r}_{m}(t)$ can be expressed in terms of the so-called lesser and greater correlations functions\cite{Haugbook}
\begin{equation}
\label{eq: lesser Green's function}
G^{<}_{m}(t)=\langle\psi^{\rm g}_{N}|\hat{a}^{\dagger}_{m}(0)\hat{a}^{}_{m}(t)|\psi^{\rm g}_{N}\rangle
\end{equation}
and
\begin{equation}
\label{eq: greater Green's function}
G^{>}_{m}(t)=\langle\psi^{\rm g}_{N}|\hat{a}^{}_{m}(t)\hat{a}^{\dagger}_{m}(0)|\psi^{\rm g}_{N}\rangle,
\end{equation}
as
\begin{equation}
\label{eq: Lz ret Green's function}
G^{\rm r}_{m}(t)=-i\Theta(t)\big[G^{>}_{m}(t)+G^{<}_{m}(t)\big].
\end{equation}
The lesser and greater correlations functions, Eqs.~\eqref{eq: lesser Green's function} and \eqref{eq: greater Green's function}, can be evaluated as follows. Explicitly  writing out the time dependence of the operators
\begin{align}
G^{<}_{m}(t)&=\langle\psi^{\rm g}_{N}|\hat{a}^{\dagger}_{m}(0)e^{i(\hat{H}-\mu\hat{N})t}\hat{a}_{m}e^{-i(\hat{H}-\mu\hat{N})t}|\psi^{\rm g}_{N}\rangle\nonumber\\&=\langle\psi^{\rm g}_{N}|\hat{a}^{\dagger}_{m}(0)e^{i(\hat{H}-\mu\hat{N})t}\hat{a}_{m}|\psi^{\rm g}_{N}\rangle e^{-i(E_{N}^{\rm g}-\mu N)t}.
\end{align}
Next, one inserts a resolution of the identity operator expressed as a complete set of  energy eigenstates of a system containing $N-1$ particles,
 \begin{align}
G^{<}_{m}(t)&=\sum_{\alpha}e^{i[E^{}_{N-1,\alpha}-\mu(N-1)]t}e^{-i(E_{N}^{\rm g}-\mu N)t}\nonumber\\&\times\langle\psi^{\rm g}_{N}|\hat{a}^{\dagger}_{m}(0)|\alpha,N-1\rangle\langle N-1,\alpha|\hat{a}_{m}|\psi^{\rm g}_{N}\rangle\nonumber\\&=\sum_{\alpha}e^{-i(E^{\rm g}_{N}-E^{}_{N-1,\alpha}-\mu)t}|\langle N-1,\alpha|\hat{a}_{m}|\psi^{\rm g}_{N}\rangle|^{2}.
\end{align}
For the  $N-1$ particle system, the monopole strength is held fixed, i.e., it corresponds to the value used to define the ground state $|\psi^{\rm g}_{N}\rangle$. 
Similarly for $G^{>}_{m}(t)$,  Eq.~\eqref{eq: greater Green's function},
\begin{equation}
G^{>}_{m}(t)=\sum_{\alpha}e^{-i(E^{}_{N+1,\alpha}-E^{\rm g}_{N}-\mu)t}|\langle N+1,\alpha|\hat{a}^{\dagger}_{m}|\psi^{\rm g}_{N}\rangle|^{2}.
\end{equation}

The Fourier transform of $G^{\rm r}_{m}(t)$, Eq.~\eqref{eq: Lz ret Green's function}, is then 
\begin{align}
\label{eq: Fourier transform of Lz resolved Green's function}
G^{\rm r}_{m}(\omega)&=\int\limits_{-\infty}^{\infty}dt\, G^{\rm r}_{m}(t)e^{i\omega t}\nonumber\\&=-i\lim_{\eta\to0^{+}}\int\frac{d\omega'}{2\pi}\frac{G^{>}_{m}(\omega')+G^{<}_{m}(\omega')}{\omega-\omega'+i\eta},
\end{align}
where
\begin{align}
G^{>}_{m}(\omega)&=2\pi\sum_{\alpha}|\langle N+1,\alpha|\hat{a}^{\dagger}_{m}|\psi^{\rm g}_{N}\rangle|^{2}\nonumber\\&\times\delta(\omega+\mu-E^{}_{N+1,\alpha}+E^{\rm g}_{N})\nonumber\\&\equiv2\pi\sum_{\alpha}Z^{>}_{m,\alpha}\delta(\omega+\mu-E^{}_{N+1,\alpha}+E^{\rm g}_{N}),
\end{align}
and
\begin{align}
G^{<}_{m}(\omega)&=2\pi \sum_{\alpha}|\langle N-1,\alpha|\hat{a}_{m}|\psi^{\rm g}_{N}\rangle|^{2}\nonumber\\&\times\delta(\omega+\mu+E^{}_{N-1,\alpha}-E^{\rm g}_{N})\nonumber\\&\equiv2\pi \sum_{\alpha} Z^{<}_{m,\alpha}\delta(\omega+\mu+E^{}_{N-1,\alpha}-E^{\rm g}_{N}).
\end{align}
The $L^{}_{z}$-resolved spectral function is then defined as 
\begin{align}
\label{eq: Lz spectral function}
A^{}_{m}(\omega)&=-\frac{1}{\pi}\,{\rm Im}\, G^{\rm r}_{m}(\omega)\nonumber\\&=\sum_{\alpha}\Big[Z^{>}_{m,\alpha}\delta(\omega+\mu-E^{}_{N+1,\alpha}+E^{\rm g}_{N})\nonumber\\&+Z^{<}_{m,\alpha}\delta(\omega+\mu+E^{}_{N-1,\alpha}-E^{\rm g}_{N})\Big],
\end{align}
which satisfies  the sum rule
\begin{equation}
\label{eq: Am sum rule}
\sum_{m=-Q}^{Q}\int\limits_{-\infty}^{\mu}d\omega\, A^{}_{m}(\omega)=N. 
\end{equation}

\begin{figure}
\includegraphics[width=\columnwidth]{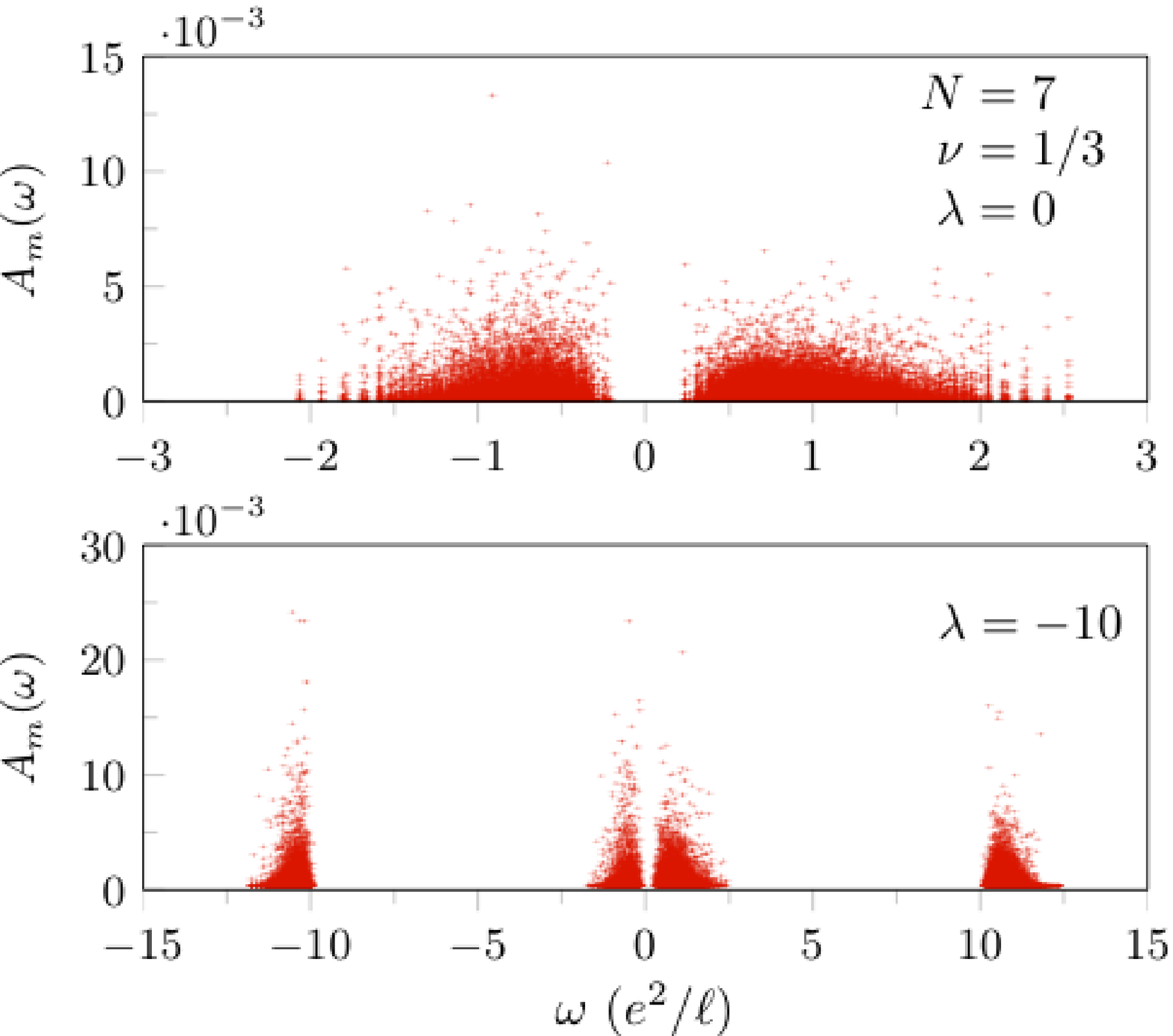}%
\caption{The top panel shows all spectral weights $Z^{\gtrless}_{m,\alpha}$ of the spectral function, Eq.~\eqref{eq: Lz spectral function}, in absence of an impurity, for $N=7$ particles at filling factor $\nu=1/3$, approximately 58,000 states. The bottom (note scale) shows the same for a system in the presence of a delta-impurity potential having strength $\lambda=-10\, e^{2}/\ell$.  \label{fig1}}
\end{figure}

\begin{figure}
\includegraphics[width=\columnwidth]{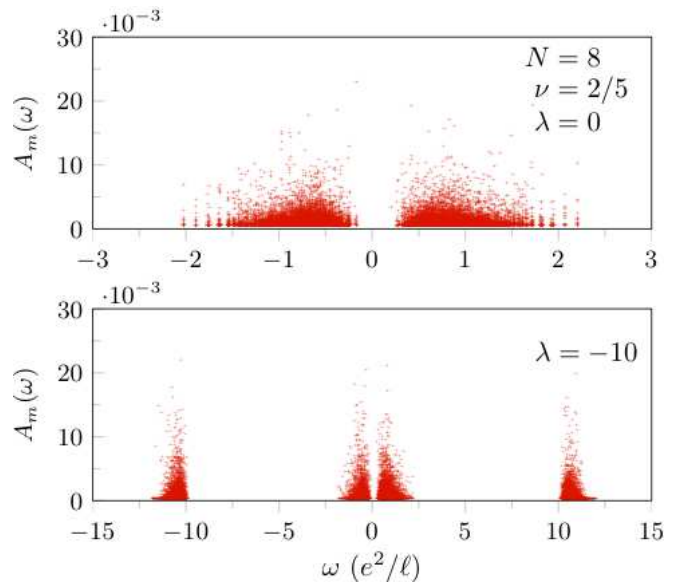}%
\caption{The top panel shows all spectral weights $Z^{\gtrless}_{m,\alpha}$ of the spectral function, Eq.~\eqref{eq: Lz spectral function}, in absence of an impurity, for $N=8$ particles at filling factor $\nu=2/5$, approximately 28,000 states. The bottom shows the same for a system in the presence of a delta-impurity potential having strength $\lambda=-10\, e^{2}/\ell$. \label{fig2}}
\end{figure}

Figures \ref{fig1} and \ref{fig2} show the calculated spectral weights $Z^{\gtrless}_{m,\alpha}$ and location of the single-particle excitation energies for $\nu=1/3$ and $\nu=2/5$ respectively,  with (bottoms panels) and without (top panels) an impurity. 
For numerical purposes the chemical potential is taken to be $\mu=(\mu_{+}+\mu_{-})/2$, where $\mu_{+}=E^{\rm G}_{N+1}-E^{G}_{N}$ and $\mu_{-}=E^{\rm g}_{N}-E^{g}_{N-1}$. For a noninteracting system in the presence of an attractive delta-impurity potential, of the form given by  Eqs.~\eqref{eq: impurity potential} and \eqref{eq: matrix elements of impurity}, a single bound state below the LLL  forms at $\omega^{}_{\rm b}\approx\omega_{\rm c}/2-\mu+\lambda\simeq\lambda$. \footnote{Technically, as was first pointed out in Ref.~\onlinecite{PrangePRB81}, to obtain a truly  localized and mathematically well-defined bound state higher Landau levels as well as a renormalization or regularization of the delta impurity have to be included.  However in  the strong field limit $\lambda/\omega_{\rm c}\ll 1$ the LLL approximate bound state energy and wave function are logarithmically accurate.}  As seen in the bottom panels of Figs.~\ref{fig1} and \ref{fig2}, for an interacting system this remains qualitatively the same. Although instead of a single bound state, many additional resonances, with an energy width of approximately $e^{2}/\ell$, appear.  As discussed in Sec. \ref{sec: T-matrix etc}, these new quasi-bound states appear because of the interaction renormalization of the bare impurity potential, effectively giving a finite width to the delta function. 

The total $L_{z}$-resolved spectral weights of the quasi-bound states are obtained by integrating the spectral functions over the energy width of the impurity resonances,
\begin{equation}
\label{eq: bound state spectral weights}
Z_{{\rm b},m}=\hspace{-0.7cm}\int\limits_{\text{impurity states}}\hspace{-0.7cm}d\omega\, A_{m}(\omega).
\end{equation}
The strength of the impurity is chosen large enough to well-separate  the quasi-bound states  from the rest of the spectrum and thus facilitate their identification and the integration region for \eqref{eq: bound state spectral weights}.  In principle they could be identified, for an arbitrary impurity strength, by the poles of the generalized $T$-matrix \eqref{eq: interacting T-matrix}, but, for finite-sized systems, this is numerically difficult.  

\section{Results \& Discussion}
\label{sec: Results}
\begin{figure}
\includegraphics[width=\columnwidth]{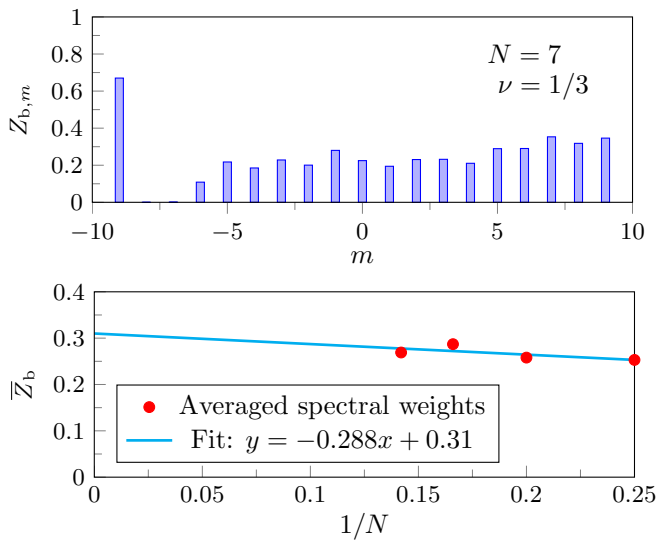}%
\caption{The top panel shows the integrated spectral weights of impurity states per angular momentum channel $m$, for $N=7$ particles at filling factor $\nu=1/3$.  The bottom panel shows the averaged integrated impurity  spectral weights verses total particle number, along with a simple linear fit to extrapolate to the thermodynamic limit.  \label{fig3}}
\end{figure}

\begin{figure}
\includegraphics{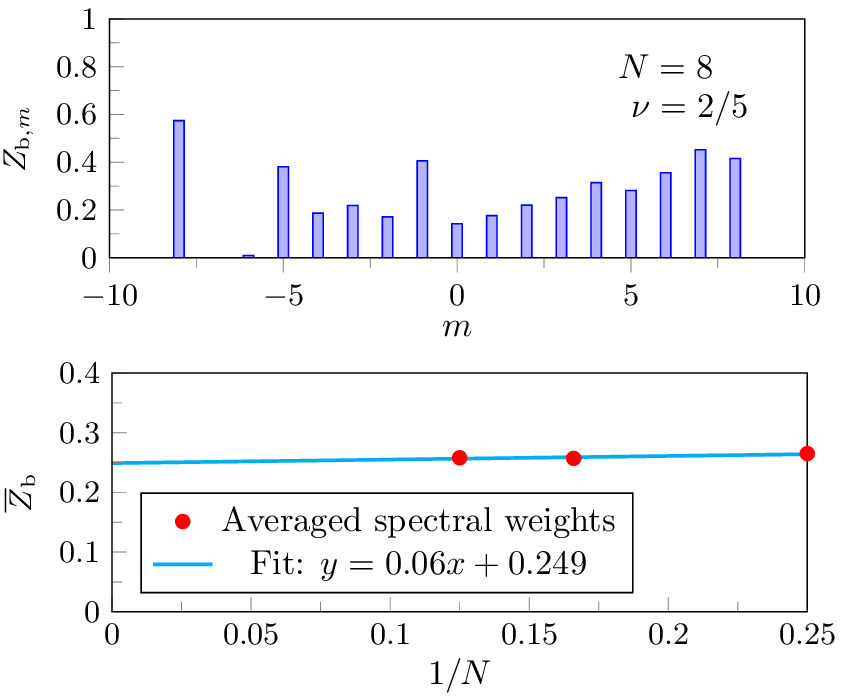}%
\caption{The top panel shows the integrated spectral weights of impurity states per angular momentum channel $m$, for $N=8$ particles at filling factor $\nu=2/5$.  The bottom panel shows the averaged integrated impurity  spectral weights verses total particle number, along with a simple linear fit to extrapolate to the thermodynamic limit. \label{fig4}}
\end{figure}

\begin{table}
\subfloat[ $\nu=1/3$]{\begin{tabular}{c c c }
\toprule
   $N$ & $\overline{Z}_{\rm b}$ & $\sigma$  \\ [0.5ex]
\hline
 4 & 0.253 & 0.265  \\ 
 5 & 0.258 & 0.198  \\ 
 6 & 0.287 & 0.152  \\ 
 7 & 0.269 & 0.117 \\ 
  [1ex]
\hline
\end{tabular}}
\hspace{1cm}
\subfloat[$\nu=2/5$]{\begin{tabular}{c c c }
\hline\hline
   $N$ & $\overline{Z}_{\rm b}$ & $\sigma$  \\ [0.5ex]
\hline
 4 & 0.265 & 0.305  \\ 
 6 & 0.257 & 0.197  \\ 
 8 & 0.258 & 0.133  \\ 
  [1ex]
\hline 
\end{tabular}}
\caption{Here, the data  shown in the bottom panels of Figs.~\ref{fig3} and Fig.~\ref{fig4} is given:  The averaged integrated impurity spectral weights $\overline{Z}_{\rm b}=(2|Q|+1)^{-1}\sum_{m}Z_{{\rm b},m}$ for each particle number $N$ and  filling factors $\nu$.  The variance $\sigma$ of the integrated impurity spectral weights is also stated. \label{table: for fig3 and fig4}}
\end{table}

For the data shown in Figs.~\ref{fig1} and \ref{fig2}, the integrated spectral weights of the impurity states  in each angular momentum channel   are shown in the top panels of Figs.~\ref{fig3} and \ref{fig4}.
The averaged  integrated spectral weights $\overline{Z}_{\rm b}=(2|Q|+1)^{-1}\sum_{m}Z_{{\rm b},m}$ of the quasi-bound states  as a function of $1/N$  are shown in the bottom panels of Figs.~\ref{fig3} and \ref{fig4} for filling factors $\nu=1/3$ and $\nu=2/5$ respectively and listed in Table \ref{table: for fig3 and fig4}.  The variance $\sigma$ of the $L_{z}$-resolved weights is also listed in Table \ref{table: for fig3 and fig4}.  While the variance is quite large, it is monotonically decreasing as the system size increases.  As can be seen from Fig.~\ref{fig3}, for $\nu=1/3$  the extrapolation to the thermodynamic limit gives a bound-state weight, or fractional charge, of approximately $e^{*}\approx0.31e$, which is consistent with the value $e^{*}=e/3$.  For $\nu=2/5$, see Fig.~\ref{fig4}, we find $e^{*}\approx0.25e$, which, although reasonable, is further from the predicted $e^{*}=e/5$ value.  The accuracy or lack of compared to the $\nu=1/3$ case could  simply be a finite-size effect, as the Hilbert space for the $\nu=1/3$ system is approximately  twice as large, even through the $\nu=2/5$ system has a greater number of electrons.  This result  could  also be due to bunching of Laughlin-quasiparticles, where two $e^{*}=e/5$ charged particles form a single quasi-bound state. This bunching effect has been recently observed in shot-noise experiments at low temperature.\cite{ChungPRL03}

Finally, we discuss what appear to be outliers, both large and small, in the data for the $L_{z}$-resolved weights. These can be most easily  seen in the top panels of Fig.~\ref{fig3} and Fig.~\ref{fig4} for the smallest $m$-values. These states are localized  in space on the opposite side of the Haldane sphere than that of the impurity potential.  For several values of angular momentum the bound-state spectral weights appear to vanish or become very small.  These correspond to extremely poor quasi-bound states with  short lifetimes.  But with no a priori reason for exclusion they are included in the average for $\overline{Z}_{\rm b}$.  
In contrast there is a single weight, at the smallest $m$-value for each filling factor, with a comparatively large integrated spectral weight, which seemingly  approaches unity.    Although this is also included in the average, as mentioned in Sec.~\ref{sec: introduction}, these states could be an indication of a bound electron-quasiparticle. Notwithstanding no other indication of such single-particle excitations  are seen in the spectral function of the impurity-free systems, e.g., an electron-quasiparticle peak at the chemical potential.  

\section{Conclusions}
In conclusion, we have calculated the fractional charge of impurity bound states in a FQH systems by an exact diagonalization calculation for the electron Green's function and extracting  the spectral weights of the impurity bound states; these correspond to the fraction of a bare electron that remains in each single-particle state.  We find  evidence consistent with the theoretically predicted fractional charge carried by Laughlin-quasiparticles: $e^{*}\approx0.31e$ for filling factor $\nu=1/3$ and $e^{*}\approx0.25e$ for $\nu=2/5$. 
\begin{acknowledgments}
KRP would like to thank Emily Pritchett for  contributions during the early stages of this work and Hartmut Hafermann for many helpful discussions and suggestions. 
\end{acknowledgments}


\begin{thebibliography}{37}%
\makeatletter
\providecommand \@ifxundefined [1]{%
 \@ifx{#1\undefined}
}%
\providecommand \@ifnum [1]{%
 \ifnum #1\expandafter \@firstoftwo
 \else \expandafter \@secondoftwo
 \fi
}%
\providecommand \@ifx [1]{%
 \ifx #1\expandafter \@firstoftwo
 \else \expandafter \@secondoftwo
 \fi
}%
\providecommand \natexlab [1]{#1}%
\providecommand \enquote  [1]{``#1''}%
\providecommand \bibnamefont  [1]{#1}%
\providecommand \bibfnamefont [1]{#1}%
\providecommand \citenamefont [1]{#1}%
\providecommand \href@noop [0]{\@secondoftwo}%
\providecommand \href [0]{\begingroup \@sanitize@url \@href}%
\providecommand \@href[1]{\@@startlink{#1}\@@href}%
\providecommand \@@href[1]{\endgroup#1\@@endlink}%
\providecommand \@sanitize@url [0]{\catcode `\\12\catcode `\$12\catcode
  `\&12\catcode `\#12\catcode `\^12\catcode `\_12\catcode `\%12\relax}%
\providecommand \@@startlink[1]{}%
\providecommand \@@endlink[0]{}%
\providecommand \url  [0]{\begingroup\@sanitize@url \@url }%
\providecommand \@url [1]{\endgroup\@href {#1}{\urlprefix }}%
\providecommand \urlprefix  [0]{URL }%
\providecommand \Eprint [0]{\href }%
\providecommand \doibase [0]{http://dx.doi.org/}%
\providecommand \selectlanguage [0]{\@gobble}%
\providecommand \bibinfo  [0]{\@secondoftwo}%
\providecommand \bibfield  [0]{\@secondoftwo}%
\providecommand \translation [1]{[#1]}%
\providecommand \BibitemOpen [0]{}%
\providecommand \bibitemStop [0]{}%
\providecommand \bibitemNoStop [0]{.\EOS\space}%
\providecommand \EOS [0]{\spacefactor3000\relax}%
\providecommand \BibitemShut  [1]{\csname bibitem#1\endcsname}%
\let\auto@bib@innerbib\@empty
\bibitem [{\citenamefont {Pan}\ \emph {et~al.}(2000)\citenamefont {Pan},
  \citenamefont {Hudson}, \citenamefont {Lang}, \citenamefont {Eisaki},
  \citenamefont {Uchida},\ and\ \citenamefont {Davis}}]{PanNature00}%
  \BibitemOpen
  \bibfield  {author} {\bibinfo {author} {\bibfnamefont {S.~H.}\ \bibnamefont
  {Pan}}, \bibinfo {author} {\bibfnamefont {E.~W.}\ \bibnamefont {Hudson}},
  \bibinfo {author} {\bibfnamefont {K.~M.}\ \bibnamefont {Lang}}, \bibinfo
  {author} {\bibfnamefont {H.}~\bibnamefont {Eisaki}}, \bibinfo {author}
  {\bibfnamefont {S.}~\bibnamefont {Uchida}}, \ and\ \bibinfo {author}
  {\bibfnamefont {J.~C.}\ \bibnamefont {Davis}},\ }\href {\doibase
  10.1038/35001534} {\bibfield  {journal} {\bibinfo  {journal} {Nature}\
  }\textbf {\bibinfo {volume} {403}},\ \bibinfo {pages} {746} (\bibinfo {year}
  {2000})}\BibitemShut {NoStop}%
\bibitem [{\citenamefont {Balatsky}\ \emph {et~al.}(2006)\citenamefont
  {Balatsky}, \citenamefont {Vekhter},\ and\ \citenamefont
  {Zhu}}]{BalastskyRMP06}%
  \BibitemOpen
  \bibfield  {author} {\bibinfo {author} {\bibfnamefont {A.~V.}\ \bibnamefont
  {Balatsky}}, \bibinfo {author} {\bibfnamefont {I.}~\bibnamefont {Vekhter}}, \
  and\ \bibinfo {author} {\bibfnamefont {J.-X.}\ \bibnamefont {Zhu}},\ }\href
  {\doibase 10.1103/RevModPhys.78.373} {\bibfield  {journal} {\bibinfo
  {journal} {Rev. Mod. Phys.}\ }\textbf {\bibinfo {volume} {78}},\ \bibinfo
  {pages} {373} (\bibinfo {year} {2006})}\BibitemShut {NoStop}%
\bibitem [{\citenamefont {Laughlin}(1983)}]{LaughlinPRL83}%
  \BibitemOpen
  \bibfield  {author} {\bibinfo {author} {\bibfnamefont {R.~B.}\ \bibnamefont
  {Laughlin}},\ }\href {\doibase 10.1103/PhysRevLett.50.1395} {\bibfield
  {journal} {\bibinfo  {journal} {Phys. Rev. Lett.}\ }\textbf {\bibinfo
  {volume} {50}},\ \bibinfo {pages} {1385} (\bibinfo {year}
  {1983})}\BibitemShut {NoStop}%
\bibitem [{\citenamefont {de~Picciotto}\ \emph {et~al.}(1997)\citenamefont
  {de~Picciotto}, \citenamefont {Reznikov}, \citenamefont {Heiblum},
  \citenamefont {Umansky}, \citenamefont {Bunin},\ and\ \citenamefont
  {Mahalu}}]{RdePicciottoNature97}%
  \BibitemOpen
  \bibfield  {author} {\bibinfo {author} {\bibfnamefont {R.}~\bibnamefont
  {de~Picciotto}}, \bibinfo {author} {\bibfnamefont {M.}~\bibnamefont
  {Reznikov}}, \bibinfo {author} {\bibfnamefont {M.}~\bibnamefont {Heiblum}},
  \bibinfo {author} {\bibfnamefont {V.}~\bibnamefont {Umansky}}, \bibinfo
  {author} {\bibfnamefont {G.}~\bibnamefont {Bunin}}, \ and\ \bibinfo {author}
  {\bibfnamefont {D.}~\bibnamefont {Mahalu}},\ }\href {\doibase 10.1038/38241}
  {\bibfield  {journal} {\bibinfo  {journal} {Nature}\ }\textbf {\bibinfo
  {volume} {389}},\ \bibinfo {pages} {162} (\bibinfo {year}
  {1997})}\BibitemShut {NoStop}%
\bibitem [{\citenamefont {Martin}\ \emph {et~al.}(2004)\citenamefont {Martin},
  \citenamefont {Ilani}, \citenamefont {Verdene}, \citenamefont {Smet},
  \citenamefont {Umansky}, \citenamefont {Mahalu}, \citenamefont {Schuh},
  \citenamefont {Abstreiter},\ and\ \citenamefont {Yacoby}}]{MartinScience04}%
  \BibitemOpen
  \bibfield  {author} {\bibinfo {author} {\bibfnamefont {J.}~\bibnamefont
  {Martin}}, \bibinfo {author} {\bibfnamefont {S.}~\bibnamefont {Ilani}},
  \bibinfo {author} {\bibfnamefont {B.}~\bibnamefont {Verdene}}, \bibinfo
  {author} {\bibfnamefont {J.}~\bibnamefont {Smet}}, \bibinfo {author}
  {\bibfnamefont {V.}~\bibnamefont {Umansky}}, \bibinfo {author} {\bibfnamefont
  {D.}~\bibnamefont {Mahalu}}, \bibinfo {author} {\bibfnamefont
  {D.}~\bibnamefont {Schuh}}, \bibinfo {author} {\bibfnamefont
  {G.}~\bibnamefont {Abstreiter}}, \ and\ \bibinfo {author} {\bibfnamefont
  {A.}~\bibnamefont {Yacoby}},\ }\href {\doibase 10.1126/science.1099950}
  {\bibfield  {journal} {\bibinfo  {journal} {Science}\ }\textbf {\bibinfo
  {volume} {305}},\ \bibinfo {pages} {980} (\bibinfo {year}
  {2004})}\BibitemShut {NoStop}%
\bibitem [{\citenamefont {Rezayi}\ and\ \citenamefont
  {Haldane}(1985)}]{RezayiPRB85}%
  \BibitemOpen
  \bibfield  {author} {\bibinfo {author} {\bibfnamefont {E.~H.}\ \bibnamefont
  {Rezayi}}\ and\ \bibinfo {author} {\bibfnamefont {F.~D.~M.}\ \bibnamefont
  {Haldane}},\ }\href {\doibase 10.1103/PhysRevB.32.6924} {\bibfield  {journal}
  {\bibinfo  {journal} {Phys. Rev. B}\ }\textbf {\bibinfo {volume} {32}},\
  \bibinfo {pages} {6924} (\bibinfo {year} {1985})}\BibitemShut {NoStop}%
\bibitem [{\citenamefont {Tsiper}(2006)}]{TsiverPRL06}%
  \BibitemOpen
  \bibfield  {author} {\bibinfo {author} {\bibfnamefont {E.~V.}\ \bibnamefont
  {Tsiper}},\ }\href {\doibase 10.1103/PhysRevLett.97.076802} {\bibfield
  {journal} {\bibinfo  {journal} {Phys. Rev. Lett.}\ }\textbf {\bibinfo
  {volume} {97}},\ \bibinfo {pages} {076802} (\bibinfo {year}
  {2006})}\BibitemShut {NoStop}%
\bibitem [{\citenamefont {Hu}\ \emph {et~al.}(2008)\citenamefont {Hu},
  \citenamefont {Wan},\ and\ \citenamefont {Schmitteckert}}]{HuPRB08}%
  \BibitemOpen
  \bibfield  {author} {\bibinfo {author} {\bibfnamefont {Z.}~\bibnamefont
  {Hu}}, \bibinfo {author} {\bibfnamefont {X.}~\bibnamefont {Wan}}, \ and\
  \bibinfo {author} {\bibfnamefont {P.}~\bibnamefont {Schmitteckert}},\
  }\href@noop {} {\bibfield  {journal} {\bibinfo  {journal} {Phys. Rev. B}\
  }\textbf {\bibinfo {volume} {77}},\ \bibinfo {pages} {075331} (\bibinfo
  {year} {2008})}\BibitemShut {NoStop}%
\bibitem [{\citenamefont {Zhang}\ \emph {et~al.}(1985)\citenamefont {Zhang},
  \citenamefont {Vulovic}, \citenamefont {Guo},\ and\ \citenamefont
  {Sarma}}]{ZhangPRB85}%
  \BibitemOpen
  \bibfield  {author} {\bibinfo {author} {\bibfnamefont {F.~C.}\ \bibnamefont
  {Zhang}}, \bibinfo {author} {\bibfnamefont {V.~Z.}\ \bibnamefont {Vulovic}},
  \bibinfo {author} {\bibfnamefont {Y.}~\bibnamefont {Guo}}, \ and\ \bibinfo
  {author} {\bibfnamefont {S.~D.}\ \bibnamefont {Sarma}},\ }\href {\doibase
  10.1103/PhysRevB.32.6920} {\bibfield  {journal} {\bibinfo  {journal} {Phys.
  Rev. B}\ }\textbf {\bibinfo {volume} {32}},\ \bibinfo {pages} {6920}
  (\bibinfo {year} {1985})}\BibitemShut {NoStop}%
\bibitem [{\citenamefont {Aristone}\ and\ \citenamefont
  {Studart}(1993)}]{AristonePRB93}%
  \BibitemOpen
  \bibfield  {author} {\bibinfo {author} {\bibfnamefont {F.}~\bibnamefont
  {Aristone}}\ and\ \bibinfo {author} {\bibfnamefont {N.}~\bibnamefont
  {Studart}},\ }\href {\doibase 10.1103/PhysRevB.47.2176} {\bibfield  {journal}
  {\bibinfo  {journal} {Phys. Rev. B}\ }\textbf {\bibinfo {volume} {47}},\
  \bibinfo {pages} {2176} (\bibinfo {year} {1993})}\BibitemShut {NoStop}%
\bibitem [{\citenamefont {V\'yborn\'y}\ \emph {et~al.}(2007)\citenamefont
  {V\'yborn\'y}, \citenamefont {M\"uller},\ and\ \citenamefont
  {Pfannkuche}}]{VybornyArxiv07}%
  \BibitemOpen
  \bibfield  {author} {\bibinfo {author} {\bibfnamefont {K.}~\bibnamefont
  {V\'yborn\'y}}, \bibinfo {author} {\bibfnamefont {C.}~\bibnamefont
  {M\"uller}}, \ and\ \bibinfo {author} {\bibfnamefont {S.}~\bibnamefont
  {Pfannkuche}},\ }\href@noop {} {\  (\bibinfo {year} {2007})},\ \Eprint
  {http://arxiv.org/abs/cond-mat/0703109} {arXiv:cond-mat/0703109} \BibitemShut
  {NoStop}%
\bibitem [{\citenamefont {Jain}(2007)}]{CompositeFermionsJainBook}%
  \BibitemOpen
  \bibfield  {author} {\bibinfo {author} {\bibfnamefont {J.~K.}\ \bibnamefont
  {Jain}},\ }\href@noop {} {\emph {\bibinfo {title} {Composite Fermions}}}\
  (\bibinfo  {publisher} {Cambridge University Press},\ \bibinfo {year}
  {2007})\BibitemShut {NoStop}%
\bibitem [{\citenamefont {Jain}\ and\ \citenamefont
  {Peterson}(2005)}]{JainPRL05}%
  \BibitemOpen
  \bibfield  {author} {\bibinfo {author} {\bibfnamefont {J.~K.}\ \bibnamefont
  {Jain}}\ and\ \bibinfo {author} {\bibfnamefont {M.~R.}\ \bibnamefont
  {Peterson}},\ }\href {\doibase 10.1103/PhysRevLett.94.186808} {\bibfield
  {journal} {\bibinfo  {journal} {Phys. Rev. Lett.}\ }\textbf {\bibinfo
  {volume} {94}},\ \bibinfo {pages} {186808} (\bibinfo {year}
  {2005})}\BibitemShut {NoStop}%
\bibitem [{\citenamefont {Conti}\ and\ \citenamefont
  {Vignale}(1998)}]{ContiJPhysCondMat98}%
  \BibitemOpen
  \bibfield  {author} {\bibinfo {author} {\bibfnamefont {S.}~\bibnamefont
  {Conti}}\ and\ \bibinfo {author} {\bibfnamefont {G.}~\bibnamefont
  {Vignale}},\ }\href {\doibase 10.1088/0953-8984/10/50/002} {\bibfield
  {journal} {\bibinfo  {journal} {J. Phys.: Condens. Matter}\ }\textbf
  {\bibinfo {volume} {10}},\ \bibinfo {pages} {L779} (\bibinfo {year}
  {1998})}\BibitemShut {NoStop}%
\bibitem [{\citenamefont {Vignale}(2006)}]{VignalePRB06}%
  \BibitemOpen
  \bibfield  {author} {\bibinfo {author} {\bibfnamefont {G.}~\bibnamefont
  {Vignale}},\ }\href {\doibase 10.1103/PhysRevB.73.073306} {\bibfield
  {journal} {\bibinfo  {journal} {Phys. Rev. B}\ }\textbf {\bibinfo {volume}
  {73}},\ \bibinfo {pages} {073306} (\bibinfo {year} {2006})}\BibitemShut
  {NoStop}%
\bibitem [{\citenamefont {Eisenstein}\ \emph {et~al.}(1992)\citenamefont
  {Eisenstein}, \citenamefont {Pfeiffer},\ and\ \citenamefont
  {West}}]{EisensteinPRL92}%
  \BibitemOpen
  \bibfield  {author} {\bibinfo {author} {\bibfnamefont {J.~P.}\ \bibnamefont
  {Eisenstein}}, \bibinfo {author} {\bibfnamefont {L.~N.}\ \bibnamefont
  {Pfeiffer}}, \ and\ \bibinfo {author} {\bibfnamefont {K.~W.}\ \bibnamefont
  {West}},\ }\href {\doibase 10.1103/PhysRevLett.69.3804} {\bibfield  {journal}
  {\bibinfo  {journal} {Phys. Rev. Lett.}\ }\textbf {\bibinfo {volume} {69}},\
  \bibinfo {pages} {3804} (\bibinfo {year} {1992})}\BibitemShut {NoStop}%
\bibitem [{\citenamefont {Boebinger}\ \emph {et~al.}(1993)\citenamefont
  {Boebinger}, \citenamefont {Levi}, \citenamefont {Passner}, \citenamefont
  {Pfeiffer},\ and\ \citenamefont {West}}]{BoebingerPRB93}%
  \BibitemOpen
  \bibfield  {author} {\bibinfo {author} {\bibfnamefont {G.~S.}\ \bibnamefont
  {Boebinger}}, \bibinfo {author} {\bibfnamefont {A.~F.~J.}\ \bibnamefont
  {Levi}}, \bibinfo {author} {\bibfnamefont {A.}~\bibnamefont {Passner}},
  \bibinfo {author} {\bibfnamefont {L.~N.}\ \bibnamefont {Pfeiffer}}, \ and\
  \bibinfo {author} {\bibfnamefont {K.~W.}\ \bibnamefont {West}},\ }\href
  {\doibase 10.1103/PhysRevB.47.16608} {\bibfield  {journal} {\bibinfo
  {journal} {Phys. Rev. B}\ }\textbf {\bibinfo {volume} {47}},\ \bibinfo
  {pages} {16608} (\bibinfo {year} {1993})}\BibitemShut {NoStop}%
\bibitem [{\citenamefont {Eisenstein}\ \emph {et~al.}(1995)\citenamefont
  {Eisenstein}, \citenamefont {Pfeiffer},\ and\ \citenamefont
  {West}}]{EisensteinPRL95}%
  \BibitemOpen
  \bibfield  {author} {\bibinfo {author} {\bibfnamefont {J.~P.}\ \bibnamefont
  {Eisenstein}}, \bibinfo {author} {\bibfnamefont {L.~N.}\ \bibnamefont
  {Pfeiffer}}, \ and\ \bibinfo {author} {\bibfnamefont {K.~W.}\ \bibnamefont
  {West}},\ }\href {\doibase 10.1103/PhysRevLett.74.1419} {\bibfield  {journal}
  {\bibinfo  {journal} {Phys. Rev. Lett.}\ }\textbf {\bibinfo {volume} {74}},\
  \bibinfo {pages} {1419} (\bibinfo {year} {1995})}\BibitemShut {NoStop}%
\bibitem [{\citenamefont {Eisenstein}\ \emph {et~al.}(2009)\citenamefont
  {Eisenstein}, \citenamefont {Pfeiffer},\ and\ \citenamefont
  {West}}]{EisensteinSolidSatcom09}%
  \BibitemOpen
  \bibfield  {author} {\bibinfo {author} {\bibfnamefont {J.~P.}\ \bibnamefont
  {Eisenstein}}, \bibinfo {author} {\bibfnamefont {L.~N.}\ \bibnamefont
  {Pfeiffer}}, \ and\ \bibinfo {author} {\bibfnamefont {K.~W.}\ \bibnamefont
  {West}},\ }\href {\doibase 10.1016/j.ssc.2009.08.004} {\bibfield  {journal}
  {\bibinfo  {journal} {Solid State Commun.}\ }\textbf {\bibinfo {volume}
  {149}},\ \bibinfo {pages} {1867} (\bibinfo {year} {2009})}\BibitemShut
  {NoStop}%
\bibitem [{\citenamefont {He}\ \emph {et~al.}(1993)\citenamefont {He},
  \citenamefont {Platzman},\ and\ \citenamefont {Halperin}}]{HePRL93}%
  \BibitemOpen
  \bibfield  {author} {\bibinfo {author} {\bibfnamefont {S.}~\bibnamefont
  {He}}, \bibinfo {author} {\bibfnamefont {P.~M.}\ \bibnamefont {Platzman}}, \
  and\ \bibinfo {author} {\bibfnamefont {B.~I.}\ \bibnamefont {Halperin}},\
  }\href {\doibase 10.1103/PhysRevLett.71.777} {\bibfield  {journal} {\bibinfo
  {journal} {Phys. Rev. Lett.}\ }\textbf {\bibinfo {volume} {71}},\ \bibinfo
  {pages} {777} (\bibinfo {year} {1993})}\BibitemShut {NoStop}%
\bibitem [{\citenamefont {Abrikosov}\ \emph {et~al.}(1975)\citenamefont
  {Abrikosov}, \citenamefont {Gorkov},\ and\ \citenamefont
  {Dzyaloshinski}}]{AGD}%
  \BibitemOpen
  \bibfield  {author} {\bibinfo {author} {\bibfnamefont {A.~A.}\ \bibnamefont
  {Abrikosov}}, \bibinfo {author} {\bibfnamefont {L.}~\bibnamefont {Gorkov}}, \
  and\ \bibinfo {author} {\bibfnamefont {I.~E.}\ \bibnamefont
  {Dzyaloshinski}},\ }\href@noop {} {\emph {\bibinfo {title} {Methods of
  Quantum Field Theory in Statistical Physics}}}\ (\bibinfo  {publisher}
  {Dover},\ \bibinfo {year} {1975})\BibitemShut {NoStop}%
\bibitem [{\citenamefont {Prange}(1981)}]{PrangePRB81}%
  \BibitemOpen
  \bibfield  {author} {\bibinfo {author} {\bibfnamefont {R.~E.}\ \bibnamefont
  {Prange}},\ }\href {\doibase 10.1103/PhysRevB.23.4802} {\bibfield  {journal}
  {\bibinfo  {journal} {Phys. Rev. B}\ }\textbf {\bibinfo {volume} {23}},\
  \bibinfo {pages} {4802} (\bibinfo {year} {1981})}\BibitemShut {NoStop}%
\bibitem [{\citenamefont {Khalilov}\ and\ \citenamefont
  {Chibirova}(2007)}]{KhalilovJPhysA07}%
  \BibitemOpen
  \bibfield  {author} {\bibinfo {author} {\bibfnamefont {V.~R.}\ \bibnamefont
  {Khalilov}}\ and\ \bibinfo {author} {\bibfnamefont {F.~K.}\ \bibnamefont
  {Chibirova}},\ }\href {\doibase 10.1088/1751-8113/40/24/013} {\bibfield
  {journal} {\bibinfo  {journal} {J. Phys. A: Math. Theor.}\ }\textbf {\bibinfo
  {volume} {40}},\ \bibinfo {pages} {6469} (\bibinfo {year}
  {2007})}\BibitemShut {NoStop}%
\bibitem [{\citenamefont {Perez}\ and\ \citenamefont
  {Coutinho}(1991)}]{FernanoAJP91}%
  \BibitemOpen
  \bibfield  {author} {\bibinfo {author} {\bibfnamefont {J.~F.}\ \bibnamefont
  {Perez}}\ and\ \bibinfo {author} {\bibfnamefont {F.~A.~B.}\ \bibnamefont
  {Coutinho}},\ }\href {\doibase 10.1119/1.16714} {\bibfield  {journal}
  {\bibinfo  {journal} {Am. J. Phys.}\ }\textbf {\bibinfo {volume} {59}},\
  \bibinfo {pages} {52} (\bibinfo {year} {1991})}\BibitemShut {NoStop}%
\bibitem [{\citenamefont {Cavalcanti}\ and\ \citenamefont
  {de~Carvalho}(1998)}]{CavalcantiJPhysAMath98}%
  \BibitemOpen
  \bibfield  {author} {\bibinfo {author} {\bibfnamefont {R.~M.}\ \bibnamefont
  {Cavalcanti}}\ and\ \bibinfo {author} {\bibfnamefont {C.~A.~A.}\ \bibnamefont
  {de~Carvalho}},\ }\href {\doibase 10.1088/0305-4470/31/10/014} {\bibfield
  {journal} {\bibinfo  {journal} {J. Phys. A: Math. Gen.}\ }\textbf {\bibinfo
  {volume} {31}},\ \bibinfo {pages} {2391} (\bibinfo {year}
  {1998})}\BibitemShut {NoStop}%
\bibitem [{\citenamefont {Ziegler}\ \emph {et~al.}(1996)\citenamefont
  {Ziegler}, \citenamefont {Poilblanc}, \citenamefont {Preuss}, \citenamefont
  {Hanke},\ and\ \citenamefont {Scalapino}}]{ZieglerPRB96}%
  \BibitemOpen
  \bibfield  {author} {\bibinfo {author} {\bibfnamefont {W.}~\bibnamefont
  {Ziegler}}, \bibinfo {author} {\bibfnamefont {D.}~\bibnamefont {Poilblanc}},
  \bibinfo {author} {\bibfnamefont {R.}~\bibnamefont {Preuss}}, \bibinfo
  {author} {\bibfnamefont {W.}~\bibnamefont {Hanke}}, \ and\ \bibinfo {author}
  {\bibfnamefont {D.~J.}\ \bibnamefont {Scalapino}},\ }\href {\doibase
  10.1103/PhysRevB.53.8704} {\bibfield  {journal} {\bibinfo  {journal} {Phys.
  Rev. B}\ }\textbf {\bibinfo {volume} {53}},\ \bibinfo {pages} {8704}
  (\bibinfo {year} {1996})}\BibitemShut {NoStop}%
\bibitem [{\citenamefont {Haldane}(1983)}]{HaldanePFL83}%
  \BibitemOpen
  \bibfield  {author} {\bibinfo {author} {\bibfnamefont {F.~D.~M.}\
  \bibnamefont {Haldane}},\ }\href {\doibase 10.1103/PhysRevLett.51.605}
  {\bibfield  {journal} {\bibinfo  {journal} {Phys. Rev. Lett.}\ }\textbf
  {\bibinfo {volume} {51}},\ \bibinfo {pages} {605} (\bibinfo {year}
  {1983})}\BibitemShut {NoStop}%
\bibitem [{Note1()}]{Note1}%
  \BibitemOpen
  \bibinfo {note} {See Ref. \protect \rev@citealpnum
  {CompositeFermionsJainBook} and references therein for further
  details.}\BibitemShut {Stop}%
\bibitem [{\citenamefont {Fano}\ \emph {et~al.}(1987)\citenamefont {Fano},
  \citenamefont {Ortolani},\ and\ \citenamefont {Colombo}}]{FanoPRB86}%
  \BibitemOpen
  \bibfield  {author} {\bibinfo {author} {\bibfnamefont {G.}~\bibnamefont
  {Fano}}, \bibinfo {author} {\bibfnamefont {F.}~\bibnamefont {Ortolani}}, \
  and\ \bibinfo {author} {\bibfnamefont {E.}~\bibnamefont {Colombo}},\ }\href
  {\doibase 10.1103/PhysRevB.34.2670} {\bibfield  {journal} {\bibinfo
  {journal} {Phys. Rev. B}\ }\textbf {\bibinfo {volume} {34}},\ \bibinfo
  {pages} {2670} (\bibinfo {year} {1987})}\BibitemShut {NoStop}%
\bibitem [{\citenamefont {Haldane}(1987)}]{QHEbook}%
  \BibitemOpen
  \bibfield  {author} {\bibinfo {author} {\bibfnamefont {F.~D.~M.}\
  \bibnamefont {Haldane}},\ }\href@noop {} {\emph {\bibinfo {title} {The
  Quantum Hall Effect}}},\ edited by\ \bibinfo {editor} {\bibfnamefont {R.~E.}\
  \bibnamefont {Prange}}\ and\ \bibinfo {editor} {\bibfnamefont {S.~M.}\
  \bibnamefont {Girvin}}\ (\bibinfo  {publisher} {Springer, New York},\
  \bibinfo {year} {1987})\ p.\ \bibinfo {pages} {303}\BibitemShut {NoStop}%
\bibitem [{\citenamefont {d'Ambrumenil}\ and\ \citenamefont
  {Morf}(1989)}]{dAmbrumenilPRB89}%
  \BibitemOpen
  \bibfield  {author} {\bibinfo {author} {\bibfnamefont {N.}~\bibnamefont
  {d'Ambrumenil}}\ and\ \bibinfo {author} {\bibfnamefont {R.}~\bibnamefont
  {Morf}},\ }\href {\doibase 10.1103/PhysRevB.40.6108} {\bibfield  {journal}
  {\bibinfo  {journal} {Phys. Rev. B}\ }\textbf {\bibinfo {volume} {40}},\
  \bibinfo {pages} {6108} (\bibinfo {year} {1989})}\BibitemShut {NoStop}%
\bibitem [{\citenamefont {Wen}\ and\ \citenamefont {Zee}(1992)}]{WenPRL92}%
  \BibitemOpen
  \bibfield  {author} {\bibinfo {author} {\bibfnamefont {X.~G.}\ \bibnamefont
  {Wen}}\ and\ \bibinfo {author} {\bibfnamefont {Z.}~\bibnamefont {Zee}},\
  }\href {\doibase 10.1103/PhysRevLett.69.953} {\bibfield  {journal} {\bibinfo
  {journal} {Phys. Rev. Lett.}\ }\textbf {\bibinfo {volume} {69}},\ \bibinfo
  {pages} {953} (\bibinfo {year} {1992})}\BibitemShut {NoStop}%
\bibitem [{\citenamefont {Banerjee}(1998)}]{BanerjeePRB98}%
  \BibitemOpen
  \bibfield  {author} {\bibinfo {author} {\bibfnamefont {D.}~\bibnamefont
  {Banerjee}},\ }\href {\doibase 10.1103/PhysRevB.58.4666} {\bibfield
  {journal} {\bibinfo  {journal} {Phys. Rev. B}\ }\textbf {\bibinfo {volume}
  {58}},\ \bibinfo {pages} {4666} (\bibinfo {year} {1998})}\BibitemShut
  {NoStop}%
\bibitem [{\citenamefont {Szabo}\ and\ \citenamefont
  {Ostlund}(1989)}]{SzaboModchem}%
  \BibitemOpen
  \bibfield  {author} {\bibinfo {author} {\bibfnamefont {A.}~\bibnamefont
  {Szabo}}\ and\ \bibinfo {author} {\bibfnamefont {N.~S.}\ \bibnamefont
  {Ostlund}},\ }\href@noop {} {\emph {\bibinfo {title} {Modern Quantum
  Chemistry: Introduction to Advanced Electronic Structure Theory}}}\ (\bibinfo
   {publisher} {McGraw-Hill, New York},\ \bibinfo {year} {1989})\BibitemShut
  {NoStop}%
\bibitem [{\citenamefont {Haug}\ and\ \citenamefont {Jauho}(1998)}]{Haugbook}%
  \BibitemOpen
  \bibfield  {author} {\bibinfo {author} {\bibfnamefont {H.}~\bibnamefont
  {Haug}}\ and\ \bibinfo {author} {\bibfnamefont {A.}~\bibnamefont {Jauho}},\
  }\href@noop {} {\emph {\bibinfo {title} {Quantum Kinetics in Transport and
  Optics of Semiconductors}}}\ (\bibinfo  {publisher} {Springer, New York},\
  \bibinfo {year} {1998})\BibitemShut {NoStop}%
\bibitem [{Note2()}]{Note2}%
  \BibitemOpen
  \bibinfo {note} {Technically, as was first pointed out in Ref.~\protect
  \rev@citealpnum {PrangePRB81}, to obtain a truly localized and mathematically
  well-defined bound state higher Landau levels as well as a renormalization or
  regularization of the delta impurity have to be included. However in the
  strong field limit $\lambda /\omega _{\protect \rm c}\ll 1$ the LLL
  approximate bound state energy and wave function are logarithmically
  accurate.}\BibitemShut {Stop}%
\bibitem [{\citenamefont {Chung}\ \emph {et~al.}(2003)\citenamefont {Chung},
  \citenamefont {Heiblum},\ and\ \citenamefont {Umansky}}]{ChungPRL03}%
  \BibitemOpen
  \bibfield  {author} {\bibinfo {author} {\bibfnamefont {Y.~C.}\ \bibnamefont
  {Chung}}, \bibinfo {author} {\bibfnamefont {M.}~\bibnamefont {Heiblum}}, \
  and\ \bibinfo {author} {\bibfnamefont {V.}~\bibnamefont {Umansky}},\ }\href
  {\doibase 10.1103/PhysRevLett.91.216804} {\bibfield  {journal} {\bibinfo
  {journal} {Phys. Rev. Lett.}\ }\textbf {\bibinfo {volume} {91}},\ \bibinfo
  {pages} {216804} (\bibinfo {year} {2003})}\BibitemShut {NoStop}%
\end{thebibliography}%

\end{document}